\documentclass[manuscript]{aastex63}
\usepackage{amsmath}
\usepackage{graphicx}
\shorttitle{Reconstructing Highly-twisted Magnetic Fields}
\shortauthors{Demcsak et al.}

\begin{document}
\title{Reconstructing Highly-twisted Magnetic Fields}
\email{victor.demcsak@sydney.edu.au}
\author{Victor M.~Demcsak}
\author{Michael S.~Wheatland}
\author{Alpha ~Mastrano}
\author{Kai E.~Yang}
\affil{Sydney Institute for Astronomy, School of Physics, The University of Sydney, Sydney, NSW 2006, Australia}

\begin{abstract}
\noindent We investigate the ability of a nonlinear force-free code to calculate highly-twisted magnetic field configurations using the Titov and D\'{e}moulin ( \textit{Astron. Astropys.} ~{\bf351}, 707, 1999) equilibrium field as a test case. The code calculates a force-free field using boundary conditions on the normal component of the field in the lower boundary, and the normal component of the current density over one polarity of the field in the lower boundary. The code can also use the current density over both polarities of the field in the lower boundary as a boundary condition. We investigate the accuracy of the reconstructions with increasing flux-rope surface twist number $N_{\textrm{t}}$, achieved by decreasing the sub-surface line current in the model. We find that the code can approximately reconstruct the Titov-D\'{e}moulin field for surface twist numbers up to $N_{\textrm{t}} \approx 8.8$. This includes configurations with bald patches. We investigate the ability to recover bald patches, and more generally identify the limitations of our method for highly-twisted fields. The results have implications for our ability to reconstruct coronal magnetic fields from observational data. 
\end{abstract}


\section{Introduction}\label{s:intro} 
A solar flare is an explosive release of magnetic energy, characterised by impulsive brightening in an active region on the Sun. The coronal magnetic field responsible for a flare cannot be measured directly, however it can be reconstructed using boundary conditions provided by vector magnetograms \citep{derosa2009critical}. Detailed knowledge of the coronal field is needed to understand the physics of solar flares and eruptions. 
\\
\\
The coronal magnetic field can be modelled as a nonlinear force free field (NLFFF). The force-free equation $(\nabla \times \vec{B}) \times \vec{B}=0$ may be rewritten
 \begin{equation}
\nabla \times \vec{B} = \alpha \vec{B},
\label{NLFFF1}
\end{equation}
where $\alpha$ is the force-free parameter. Taking the divergence of Equation \ref{NLFFF1} and applying the divergence-free condition 
\begin{equation}
\nabla \cdot \vec{B}=0
\label{NLFFF2}
\end{equation}
gives 
\begin{equation}
\vec{B}\cdot \nabla \alpha=0,
\label{NLFFF3}
\end{equation}
which means that the gradient of $\alpha$ is everywhere perpendicular to the magnetic field. 
\\
\\
Many numerical methods to solve the NLFFF equations have been developed. There are three main types. The first is the Grad-Rubin approach \citep{grad1958hydromagnetic}, which has been implemented by \cite{amari1999iterative}, \cite{Wheatland_2007} and \cite{Amari_2006}. The second is the magnetofrictional method \citep{Valori_MHD_test_tD_model}, and the third is the optimisation method  \citep{Wheat_optimize_2010, Wiegelman_2010, Wiegel_2012}. A hybrid optimisation/Grad-Rubin approach appears in \cite{Amari_aly_2010}. This paper uses the CFIT code, which implements a Grad-Rubin scheme as described in \cite{Wheatland_2007}. In this approach linearised versions of Equations \ref{NLFFF1} and \ref{NLFFF3} are iteratively solved for the magnetic field ($\vec{B}$) and the force-free parameter ($\alpha$). The code works with a vector potential to ensure that Equation \ref{NLFFF2} is satisfied. 
\\ 
\\
In principle NLFFF methods can be used to reconstruct solar coronal magnetic fields using photospheric vector fields. However in practice there are many difficulties with this procedure. \cite{schrijver2006nonlinear} and \cite{metcalf2008nonlinear} investigated the ability of different NLFFF algorithms to reconstruct an analytic NLFFF \citep{low1990modeling}, as well as to reconstruct a more realistic but still synthetic test case involving a non-force-free lower boundary. Whilst the methods produced relatively accurate reconstructions of the \cite{low1990modeling} test case, they were less successful with the field with the forced lower boundary. There are also examples in the literature of NLFFF reconstructions for particular active regions where different investigators present quantitatively different results (\textit{e.g.} see reconstructions of AR 12158 by \cite{zhao2016hooked}; \cite{vemareddy2016sunspot} and \cite{lee2018mhd}). 
\\
\\
Reconstructions are often used to investigate the topology of coronal magnetic fields, and in particular topological structures which may be involved in magnetic energy release. A bald patch is a region at the photosphere where overlying magnetic field lines are tangent to the photosphere \citep{titov1993conditions}. Bald patches are the precursors for the emergence of null points into the solar coronal field \citep{bungey_bp_96}, which are preferential sites for energy release. Bald patches may also play a role in reconnection in the absence of nulls, with the energy release occurring along the separatrix surface associated with the bald patch \citep{demoulin2006extending, Titov_demoulin_99}. Bald patches can be identified in vector magnetograms, but they may not be present in NLFFF reconstructions from the data. A discussion of this point appears in \cite{derosa2009critical}. The ability to reconstruct bald patches, especially in highly twisted magnetic fields is desirable to accurately investigate the topology of the magnetic field in the corona.
\\
\\
Reconstruction methods may be tested using known magnetic field models. The Titov-D{\'e}moulin field \citep{Titov_demoulin_99} hereafter TD99, is {an analytic}, approximately force-free field, intended to model a magnetic flux rope in the corona. This field consists of a toroidal flux rope that carries a current $I$, together with a pair of sub-photospheric magnetic charges and a line current $I_{0}$, along the sub-photospheric axis of symmetry of the torus. These field components produce a force-free equilibrium for points along the circular axis of the torus. The plane which represents the photosphere, is the lower boundary of the computational domain. The field in the computational domain is the vector sum of a potential field due to the sub-photospheric charges ${\vec{B}}_{q}$, a field ${\vec{B}}_{I}$ due to the uniformly distributed current $I$ in the torus, and a potential field ${\vec{B}}_{\theta}$ due to the line current $I_{0}$. The surface twist number of the flux tube is denoted as $N_{\textrm{t}}$. This is the number of turns of a field line at the edge of the flux rope around the axis, over the whole length of the torus. The TD99 field is force free along the circular axis of the flux rope and outside the flux rope (where the field is potential). It is approximately force free at points inside the flux rope away from the flux rope axis.
\\
\\
The Titov-D{\'e}moulin field has been used previously as a test for nonlinear force-free methods. \cite{Valori_MHD_test_tD_model} applied a magnetofrictional extrapolation code to a modified version of the TD99 field where the line current from the TD99 model was replaced by a dipole. They produced an accurate reconstruction of the field for flux ropes with moderate twist. \cite{Valori_MHD_test_tD_model} characterised the twist using the average angle $\left< \Phi \right>$ through which a field line turns around the axis of the flux rope over the part of the torus in the computational domain. To estimate the corresponding surface twist numbers for the whole length of the torus, we can use the approximation:
\begin{equation}
N_{\textrm{t}} \approx \frac{ \Phi_{\textrm{t}} }{2 \cos^{-1}{\left(\frac{d}{R}\right)} },
\label{valori_aprox_twist}
\end{equation}
where $R$ is the major radius of the torus, and $\Phi_{\textrm{t}}$ is the surface twist angle over the part of the flux rope in the computational domain corresponding to the surface twist number $N_{\textrm{t}}$ over the whole torus. To derive Equation 4, notice that the length of the flux rope above the photosphere is approximately $2R\cos^{-1}(d/R)$ assuming a thin flux rope \citep{torok2004ideal}. The total length of the torus is $2\pi R$, so it follows that the surface twist $N_{\textrm{t}}$ in the whole torus is:
$$N_{\textrm{t}} \approx \frac{2\pi R}{2R\cos^{-1}(d/R)} \cdot \frac{\Phi_{\textrm{t}}}{2\pi} =\frac{\Phi_{\textrm{t}}}{2\cos^{-1}(d/R)}.$$
If $N_{\textrm{t}}$ and $d/R$ are known, then Equation 4 can be rearranged to calculate $\Phi_{\textrm{t}}$. For the TD99 model the twist in the flux rope is largest at the edge of the flux rope, so we have $\left< \Phi \right> < \Phi_{\textrm{t}}$. {\cite{Valori_MHD_test_tD_model} considered three different twists ($\left< \Phi \right >$ = $1.8 \pi$, $2.1 \pi$, and $2.7\pi$)}. It is to be noted that \cite{Valori_MHD_test_tD_model} used a parabolic radial profile for the axial current, which is different from the approximately uniform distribution of current that was used in TD99.
\\
\\
The magnetofrictional method may encounter difficulties in reconstructing a field with a highly twisted flux rope from its boundary conditions, since the method can be subject to the same instabilities as in magnetohydrodynamics (MHD). The relevant instabilities are the helical kink \citep{hood1979kink} and the torus \citep{kliem2006torus} instabilities. In particular the kink instability has been shown to occur in an MHD simulation initiated with the TD99 equilibrium once a critical average twist angle is exceeded \citep{torok2004ideal}.
\\
\\
Other reconstructions of the TD99 field have also been reported. \cite{Wiegel_TD_use_optimisation} applied the optimisation method. They used a twist number $N_{\textrm{t}} \approx 2.6$, and investigated two main cases, one where the boundary conditions were specified on all six faces of the rectangular computational domain; and a second other where only the boundary conditions for the lower boundary were supplied. The case with complete boundary information produced better results. The TD99 field has also been investigated using magnetofrictional \citep{guo2016magneto} and MHD extrapolation \citep{jiang2016testing} techniques. In both cases, the parameters used matched those in \cite{Valori_MHD_test_tD_model}.  In each case the methods succeeded in reproducing features of the TD99 model, including bald patches, for low/moderately twisted systems.  
\\
\\
Previous reconstructions of the TD99 field have {worked with modifications of the original field}. \cite{Valori_MHD_test_tD_model} modified the geometry and topology of their test field in three ways. One modification involved changing the radial profile of the axial current from uniform to parabolic. Another modification was to replace the line current by sub-surface dipoles that were oriented almost vertically, where the orientation was chosen to match the torus geometry as closely as possible. This makes the TD model more realistic, as coronal fields do not decrease with distance like the field from a line current, as in the original TD99 field. To produce boundary conditions with a bald patch \cite{Valori_MHD_test_tD_model} altered the aspect ratio $(R/a)$ of the torus. In order to apply magnetofrictional extrapolation techniques, \cite{Valori_MHD_test_tD_model} first applied a MHD relaxation to the TD99 field. This was done because the TD99 field is not strictly force free, and in particular the boundary conditions are inconsistent with a force-free field. The magnetofrictional method preserves the vector field in the lower boundary, and the initial inconsistency means that it will not produce an exactly force-free field. Also, discretisation errors imply initial $\vec{J} \times \vec{B}$ forces in the TD99 field on a numerical grid.
\\
\\
The above discussion identifies a limitation of magnetofrictional and MHD extrapolation techniques \citep{guo2016magneto, jiang2016testing, Valori_MHD_test_tD_model} to reconstruction. It is necessary to avoid fields with high twist, to prevent the kink instability, and this makes it difficult to investigate changes in topology with increasing twist number. In principle the Grad-Rubin approach can be applied to fields with high values of $N_{\textrm{t}}$. In this paper we reconstruct the TD99 field with a range of surface twist numbers obtained by decreasing the line current $I_{0}$, whilst keeping the geometric parameters and the values of the sub-photospheric charges unchanged.
\\
\\
In this paper we show that the application of a Grad-Rubin type NLFFF code (CFIT) to the TD99 field can reconstruct the main features of the field, for high surface twist numbers $N_{\textrm{t}}$. The reconstructed fields include bald patches. The TD99 parameters are chosen so that a bald patch will appear provided the twist is sufficiently high. 
\\
\\
This paper is organised as follows. Section 2 describes the parameters for the TD99 field and explains the methods used in the NLFFF reconstructions. In Section 3 we present the results of the calculations, and give an analysis of the effect of increasing surface twist number on the success of the reconstructions. In Section 4 we summarise the conclusions.
\section{Methods}\label{s:methods} 
In this section we describe how CFIT works and explain the boundary conditions for the solution cube. We identify our choice of parameters and explain how it relates to the magnetic topology we subsequently investigate. We briefly review how the TD99 field is calculated, and how the boundary conditions for the reconstructions are chosen.
\subsection{The NLFFF Code (CFIT)}
The ``current-field iteration'' or CFIT code \citep{Wheatland_2007} is an implementation of the method discussed in \cite{grad1958hydromagnetic}. There are two steps at each iteration. First, for each point in the volume, an $\alpha$ value is assigned based on propagation of boundary values of $\alpha$ along the field line threading the point. There are two choices for the assignment of $\alpha$: either the positive, or negative polarity end of the field line (for field lines anchored in the lower boundary at both ends). Second, a Fourier transform approach is used to calculate an updated field in the computational volume due to the currents implied by the $\alpha$ values in the volume and the field. The iteration sequence is started with a potential field, and continued until the field in the volume is unchanging with iteration. Since this approach uses the vector potential (instead of $\vec{B}$) the divergence-free condition is preserved up to truncation error. 
\\
\\
Because there are two choices of polarity for the boundary values of $\alpha$ at the lower boundary, there are two possible solutions, the P and N solutions. These correspond to the choice of the positive or negative polarity, respectively. In addition, a third solution (the self-consistent solution) can be constructed by averaging the boundary values of $\alpha$ in the P and N solutions, constructing new P and N solutions, and repeating this process until the result is unchanged \citep{wheatland2009self}. The self consistency procedure may be useful when the P and N solutions are different, which can be the case for vector magnetogram data \citep{derosa2009critical}. 

\subsection{The Titov-D{\'e}moulin Test Field}
The Titov-D{\'e}moulin field \citep{Titov_demoulin_99} is an axisymmetric magnetic field which is approximately force free. The Lorentz force due to a toroidal flux rope current and the magnetic field of sub-photospheric magnetic charges is approximately balanced by the self-force resulting from the flux-rope curvature. The torus is force free along its circular axis, and approximately force free away from the axis. The minor radius of the torus is required to be much less than the major radius for the approximation to be good. 
\\
\\
The parameters of the model are: $I_{0}$, the line current, $R$, the major radius of  the torus, $a$, the minor radius of the torus, $L$, the monopole distance along the flux rope axis of symmetry, $d$, the depth below the bottom of the box of the axis of symmetry, $q$, the magnetic charge, and $I$, the ring current. There are six independent parameters ($I_{0}, R, a, L, d, q$). The number of turns at the surface of the flux rope, or surface twist number $N_{\textrm{t}}$, is given by
\begin{equation}
N_{\textrm{t}} = \frac{I}{|I_{0}|} \left(\frac{R}{a} \right)^{2}.
\label{twisteqtn5}
\end{equation}

\vspace{5mm}

\noindent The parameter choices for the modelling are summarised in Table 1. We consider five cases with increasing surface twist number. The geometric parameters ($R,a,L,d$) as well as the value of the charge $q$ are fixed, which implies the ring current $I$ is also fixed. Varying $I_{0}$ changes $N_{\textrm{t}}$ according to Equation \ref{twisteqtn5}.
This allows one to increase the surface twist number whilst maintaining the same geometry. Intuitively, what we are doing is increasing the axial field in the flux rope, whilst keeping the azimuthal field in the flux rope constant.  \cite{Titov_demoulin_99} discuss how to estimate the value of $N_{\textrm{t}}$ required to produce a bald patch, and based on this estimate, Cases 2-5 are expected to contain a bald patch. Our parameter choice enables us to  explore the ability of CFIT to reconstruct fields with different magnetic topologies. We are interested in determining the highest surface twist number field which can be successfully reconstructed.
\\
\begin{table}
\centering
\begin{tabular}{l*{8}{c}r}
Field \hspace{-3mm}& $I_{0}$ $[\textrm{TA}]$  & $N_{\textrm{t}}$ &  $R$ $[\textrm{Mm}]$ & $a$ $[\textrm{Mm}]$ & $L$ $[\textrm{Mm}]$ & $d$ $[\textrm{Mm}]$ &$ q$ $[\textrm{T Mm}^{2}]$ &  ${\textrm{BP}} $ \\
\hline
Case 1 & 13.0 & 2.0 & 100 & 35 &60 & 60 & 80&  N \\
\hline
Case 2 & 8.3 & 3.2 & 100 & 35 &60 & 60 & 80&  Y \\
\hline
Case 3 & 6.0 & 4.4 & 100 & 35 &60 & 60 & 80&  Y \\
\hline 
Case 4 & 3.3 & 8.0 & 100 & 35 &60 & 60 & 80&  Y \\
\hline 
Case 5 & 3.0 & 8.8 & 100 & 35 &60 & 60 & 80&Y   \\
\hline 
\end{tabular}
\caption{Selected parameters for the TD99 equilibrium field. The five cases include {four} fields with a bald patch (BP).}
\end{table}

\vspace{-3mm}

\subsection{Boundary Conditions for the Reconstructions}
CFIT works in a cartesian domain. The boundary conditions for the construction of P and N solutions are the values of $B_{z}$ in the lower boundary together with the values of $\alpha$ in the lower boundary over the chosen sign (polarity) of $B_{z}$. The side boundaries have periodic boundary conditions. The top boundary is closed, \textit{i.e.} $B_{z}=0$ at the top of the domain. This boundary condition, together with the periodic side boundary conditions, implies that the lower boundary must be flux-balanced. This condition is true for our calculations.
\\
\\
The exact TD99 field is constructed using a separate code. Values of $B_{z}$ and $J_{z}$  are extracted from the lower boundary of the TD99 solution cube created with the separate code, and the force free parameter is calculated \textit{via}
\begin{equation}
\alpha =\left. \frac{\mu_{0} J_{z}}{B_{z}}\right |_{z=0},
\label{666y}
\end{equation}
for points with $J_{z}>0$ (boundary points inside the flux rope). For points outside the flux rope we take $\alpha=0$. The resulting values of $\alpha$ and $B_{z}$ are used to produce the P and N solution boundary conditions for the CFIT code. The size of the Titov-D{\'e}moulin boundary region is $-600 \textrm{ Mm} \le x \le 600  \textrm{ Mm}$, $-600 \textrm{ Mm} \le y \le 600  \textrm{ Mm}$, represented on a 500 $\times$ 500 grid. The reconstruction is performed with this boundary and with a choice of 85 points in the vertical direction. This volume is chosen to ensure that field lines with $\alpha \neq 0$ connect to the lower boundary at both ends, and boundary effects are relatively unimportant \citep{Valori_MHD_test_tD_model}.

\subsubsection{Instabilities}
\cite{Titov_demoulin_99} present an expression for the variation of the net force at the centre of the flux rope as a function of a variation $\delta R$ in the radius of the flux rope. If the net force variation is positive for positive $\delta R$, the flux rope is unstable in a dynamical evolution (the torus instability). \cite{Titov_demoulin_99} showed that this condition leads to a threshold $R \gtrsim \sqrt{2} L$ for the configuration to be torus unstable. For our cases $R=100$ Mm $> \sqrt{2}L \approx 85$ Mm, which suggests the configurations are unstable. However, \cite{torok2004ideal} found stable TD99 equilibria using MHD simulations for the parameters $L$=50 Mm and $R$=110 Mm, which further exceeds the TD99 threshold. This suggests that all of the cases in Table 1 are torus-stable.
\\
\\
\cite{torok2004ideal} discuss the kink instability for a dynamical evolution starting from the TD99 field. The onset of the kink instability depends on the twist and the aspect ratio of the torus. \cite{torok2004ideal} found that for an aspect ratio $R/a \approx 5$ there exists a critical average angle twist $\left< \Phi \right> \approx 3.5\pi$, which if exceeded leads to a kink instability. Note that $\left< \Phi \right>$ denotes the average angle a field line turns around the axis of the rope in the computational domain. It was found that the critical average angle twist $\left< \Phi \right>$ increases as the aspect ratio increases and vice versa. The parameters in Table 1 are different from those discussed in \cite{torok2004ideal}, thus a direct comparison is not possible. In particular the aspect ratio of our parameter choice (Table 1) is $R/a \approx 2.9$. Using Equation 4 {with the parameters in Table 1, we obtain $\Phi_{\textrm{t}}=1.8\pi$ for Case 1 ($N_{\textrm{t}} = 2.0$) and $\Phi_{\textrm{t}}=5.2\pi$ for Case 5 ($N_{\textrm{t}} = 8.8$). Hence we can expect $\left< \Phi \right> < 1.8\pi$ for Case 1, and $\left< \Phi \right> < 5.2\pi$ for Case 5. A more detailed investigation would be needed to determine whether our higher-twist cases are likely to be kink unstable.}
\subsection{Analysis of Reconstructions} \label{s:AoR} 
We investigate the success of the reconstructions of the TD99 fields in different ways. To identify the presence of relevant topological features we visually inspect field line traces for the calculated magnetic fields. We also qualitatively compare the location of bald patches along the neutral line, and the distribution of the current density $J_{z}$ at the lower boundary. The results are presented in Section 3.
\\
\\
The appearance of bald patches is dependent on the flux rope surface twist number. A bald patch exists at a polarity inversion line (or neutral line) if 
\begin{equation}
\left. \vec {B}_{\rm{h}} \cdot \vec {\bf{\nabla}}_{\rm{h}} B_{z}\right |_{\tiny{z=0}} >0,
\label{bpequation}
\end{equation}
where the subscript $\rm{h}$ indicates the horizontal component of the magnetic field, $\vec{B}_{\textrm{h}}=(B_{x}, B_{y}, 0)$, and the inequality is evaluated at the polarity inversion (neutral) line. Equation \ref{bpequation} is used to identify bald patches. 
\\
\\
The metrics presented in \cite{schrijver2006nonlinear} and \cite{metcalf2008nonlinear} are used to quantitatively compare the analytic and the reconstructed fields. Specifically, we consider the vector correlation ($C_{\textrm{vec}}$), Cauchy-Schwartz correlation ($C_{\textrm{cs}}$), relative magnetic energy ($\varepsilon $) and the mean vector errors ($E_{\textrm{n}}'$ and $E_{\textrm{m}}'$) defined respectively by:

\begin{equation}
C_{\scriptsize{\textrm{vec}}} = \frac{\sum_{i=1}^{k} \vec{B}_{i} \cdot \vec{b}_{i}}{\sqrt{\sum_{i=1}^{k} |\vec{B}_{i}|^{2}} \sqrt{\sum_{i=1}^{k} |\vec{b}_{i}|^{2}} },
\label{em1}
\end{equation}
\begin{equation}
C_{\scriptsize{\textrm{cs}}} = \frac{1}{k} \sum_{i=1}^{k} \frac{\vec{B}_{i} \cdot \vec{b}_{i}}{|\vec{B}_{i}||\vec{b}_{i}|},
\end{equation}
\begin{equation}
E_{\scriptsize{\textrm{n}}}' = 1 - \frac{\sum_{i=1}^{k} |\vec{B}_{i} - \vec{b}_{i}|}{\sum_{i=1}^{k} |\vec{B}_{i}|},
\end{equation}
\begin{equation}
E_{\scriptsize{\textrm{m}}}' = 1-\frac{1}{k} \sum_{i=1}^{k} \frac{|\vec{B}_{i} - \vec{b}_{i}|}{|\vec{B}_{i}|},
\end{equation}
and
\begin{equation}
\epsilon = \frac{\sum_{i=1}^{k} \vec{b}_{i} \cdot \vec{b}_{i}}{\sum_{i=1}^{k} \vec{B}_{i} \cdot \vec{B}_{i}},
\label{emlast}
\end{equation}

\noindent where the test field is $\vec{B}_{i}$ and the reconstructed field is $\vec{b}_{i}$. For a perfect reconstruction each metric equals unity. Note that we are reconstructing an approximately force-free magnetic field using a method which produces a strictly force-free field, so we do not expect to achieve {a perfect reconstruction}.
\\
\\
The CFIT solution is compared with the TD99 test field by calculating the metrics for a restricted analysis volume of size [66, 66, 30] in grid points, equivalent to [158.4, 158.4, 102] in Mm. This restricted size is significantly smaller than the computational volume, has as lower boundary that is part of the lower boundary of the complete volume, and is centred on the flux rope. Our restricted analysis volume contains the  features of interest, and in particular the vertical height is chosen such that it is clear of the top of the flux-rope torus. This procedure of restricting the analysis volume follows earlier studies \citep{Valori_MHD_test_tD_model, schrijver2006nonlinear,metcalf2008nonlinear}.

\section{Results}\label{s:results} 
In this section we present the results of reconstructions for the cases in Table 1, which span twist numbers $N_{\textrm{t}} = 2.0$ to $N_{\textrm{t}} = 8.8$.

\subsection{Field Lines and Current Density at the Boundary}\label{31} 


\begin{figure}
\centering
\includegraphics[width=1\textwidth]{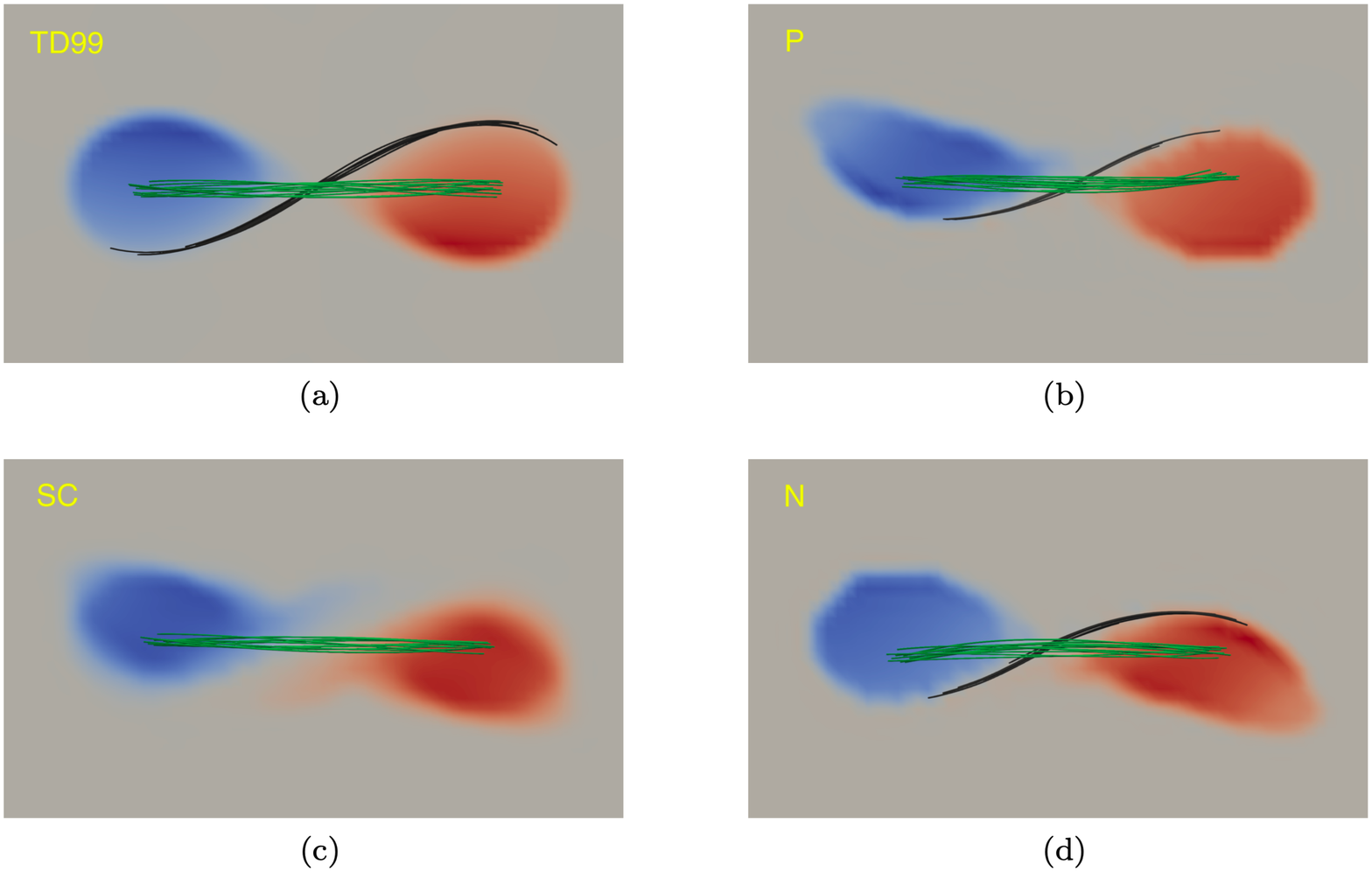}
 \caption{Flux tube system corresponding to Case 2 ($N_{\textrm{t}}=3.2$). The {red and blue colours show} regions in the lower boundary with positive and negative $J_{z}$, respectively. The green and black lines are field lines close to the centre of the flux rope, and originating in the bald patch, respectively. The panels are arranged as follows: (a) analytic test field; (b) P solutions; (c) self-consistent solution; (d) N solution.}
\end{figure}


\noindent  We investigate in detail Cases 2, 3 and 5, which correspond to Figures 1, 2 and 3 respectively. 

\begin{figure}
\centering
\includegraphics[width=1\textwidth]{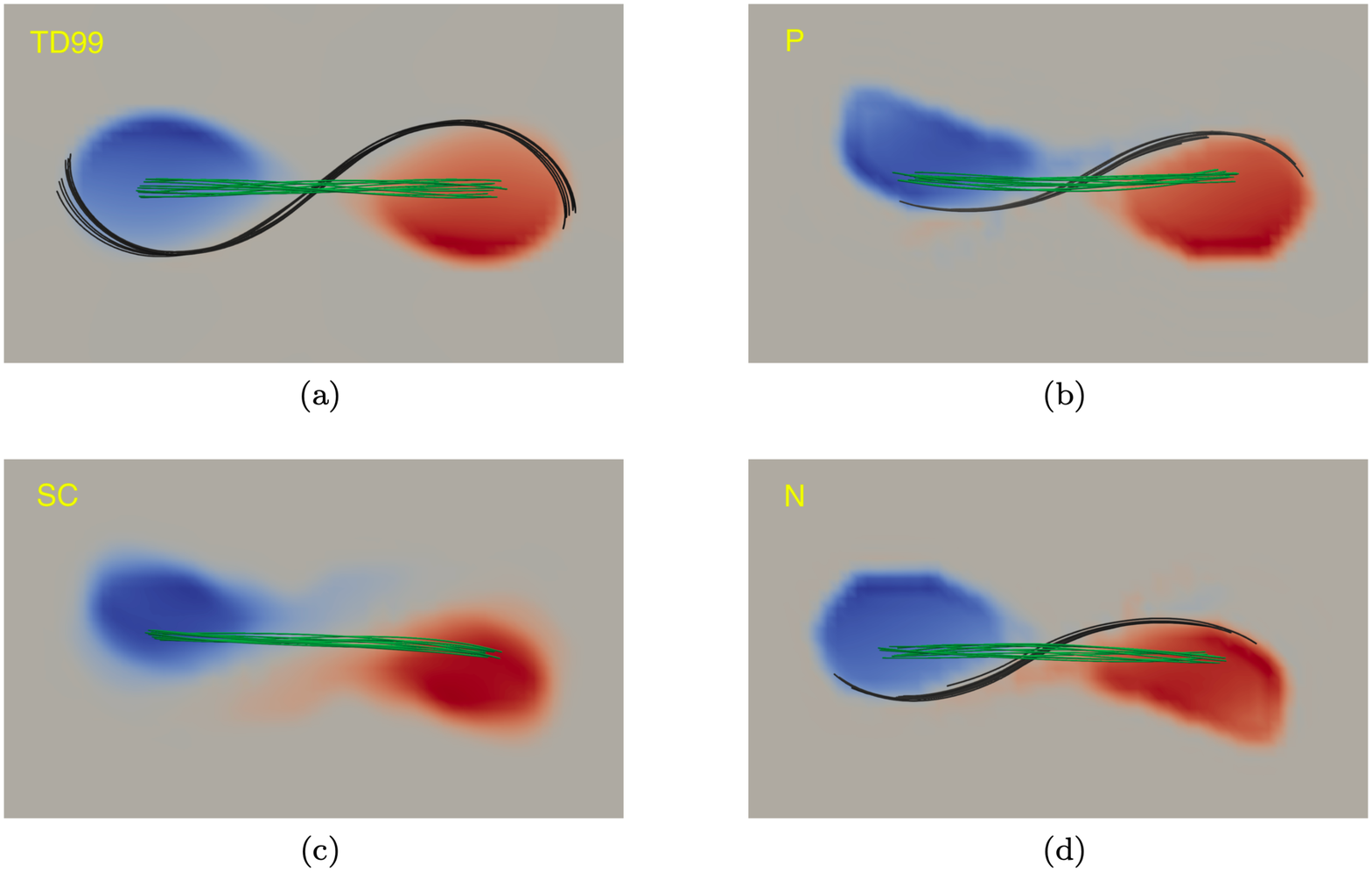}
 \caption{Flux tube system corresponding to Case 3 ($N_{\textrm{t}}=4.4$). The layout is the same as in Figure 1.}
\end{figure}

\noindent 
Figure 1 shows Case 2, with a surface twist number $N_{\textrm{t}}= 3.2$. The four panels represent the exact TD99 model, and the three reconstructions: the P solution, the N solution, and the self-consistent solution. In each panel the perspective is looking down on the computational volume, centred on the flux rope. For this case the TD99 model contains a bald patch. The black lines are magnetic field lines at the edge of the flux rope which touch the lower boundary at the bald patch, and the green lines correspond to magnetic field lines at the centre of the flux rope, winding around its axis. The axis is determined by the location in the plane $y=0$ where both $B_{x}$ and $B_{z}$ are zero, so $B_{h} = \sqrt{{B_{x}}^{2}+{B_{z}}^{2}}=0$. The red and blue colours show regions in the lower boundary with positive and negative $J_{z}$ respectively.
\\
\\
The central part of the flux rope (green) maps between the positive and negative $J_{z}$ footpoints in a similar way to the TD99 test case. The P and N solutions [panels (b) and (d)] achieve the same result, {with an interchange in the roles of the polarities. Specifically, comparing the P and N solutions [panels (b) and (d)] we see that the pattern of current density is the same after a rotation of $180^{\circ}$ and a change in the sign of the current.} This symmetry is a consequence of the CFIT solution procedure. {The code preserves the values of $B_{z}$ in the lower boundary, and the values of $\alpha$ on the chosen polarity (the positive polarity for the P solution, and the negative polarity for the N solution). Hence the code preserves $J_{z} = \alpha B_{z}$ on the chosen polarity assuming a force-free field. This may be seen by comparing panels (b) and (d) with panel (a). The current density on the other polarity is determined by the connectivity of field lines in the solution. }
The P and N solutions have the symmetry shown because of the symmetry in the boundary conditions used, \textit{i.e.} in the TD99 field the values of $B_{z}$ in the boundary conditions for the P and N solutions change sign under a $180^{\circ}$ rotation, and the values of $\alpha$ (calculated using the method described in Section 2.3) are the same under a $180^{\circ}$ rotation. For the self consistent solution [panel (c)], the averaging of the values of $\alpha$ during the construction of the solution leads to the symmetry in the distribution of $J_{z}$ shown. 
\\
\\
For Case 2, the field lines that touch the bald patch in the TD99 model form a moderately curved bald patch field line as shown in Figure 1a. The specific location of the bald patch can be determined using Equation \ref{bpequation}, and this is discussed further in Section 3.2. The P and N solutions in Figures 1b and 1d approximately reproduce the shape and location of the S-shaped field lines seen in the TD99 model. The self-consistency procedure does not reconstruct a bald patch, so the black field lines are missing from Figure 1c. For the parameters chosen in this paper, the self-consistent solution does not provide an improved reconstruction compared with the P and N solutions. This is true for all cases in this paper.

\begin{figure}
\centering
\includegraphics[width=1\textwidth]{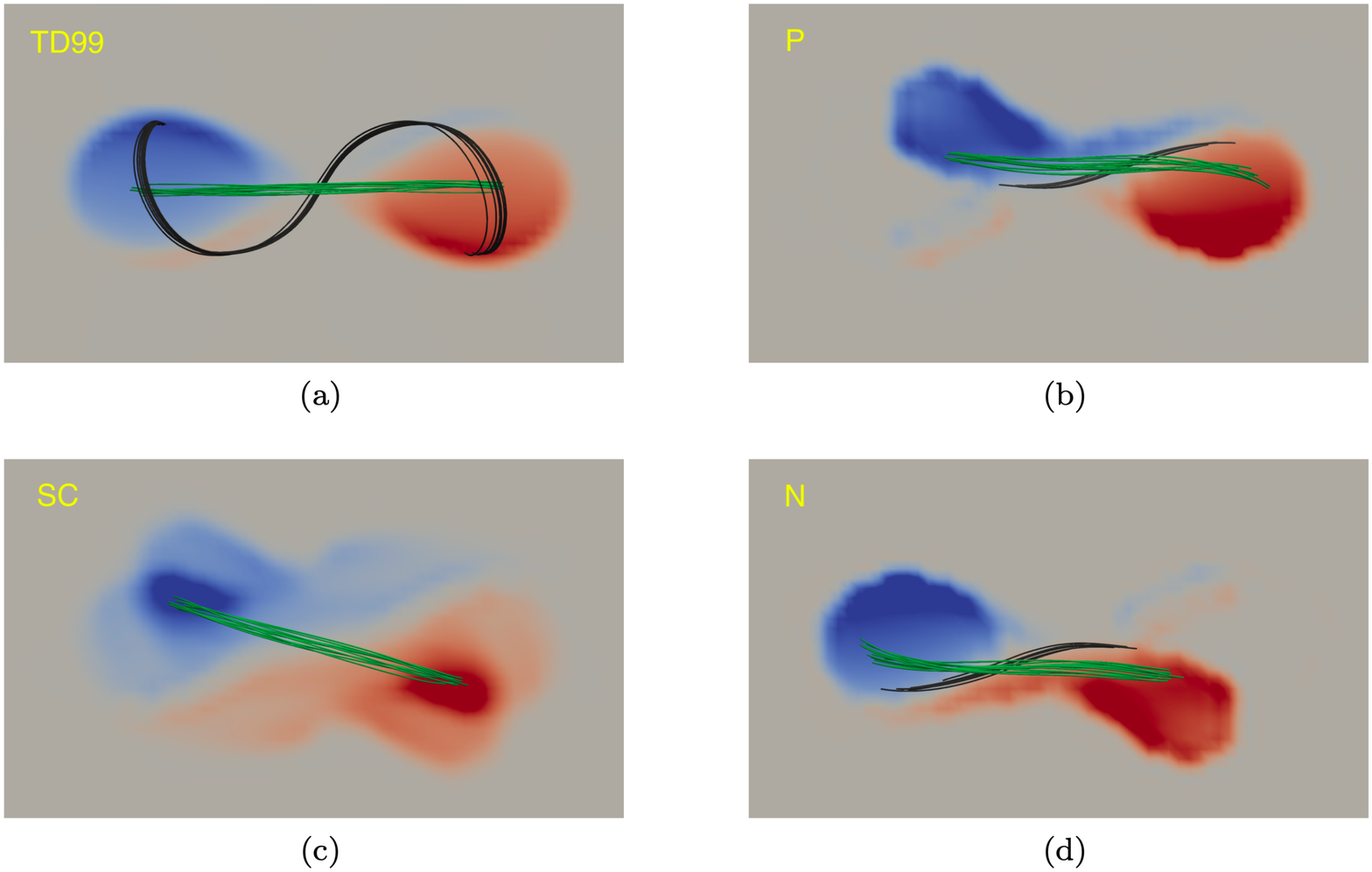}
  \caption{Flux tube system corresponding to Case 5 ($N_{\textrm{t}}=8.8$). The layout is the same as in Figure 1.}
\end{figure}

\noindent Figure 2 shows the results for Case 3, with a surface twist number $N_{\textrm{t}}=4.4$. The layout is the same as in Figure 1. In this case the TD99 test field has curved S shaped bald patch field lines, as shown in Figure 2a. The field lines near the centre of the flux rope (green) are approximately reproduced in each reconstruction. The bald patch field lines are approximately reconstructed in the P and N solutions (black curves in Figures 2b and 2d respectively). As in Case 2, no bald patch field lines are present in the self-consistent solution.
\\
\\
Figure 2 also shows that the patterns of current density $J_{z}$ in the lower boundary are recovered in the reconstructions for Case 3. The expected symmetry between the P and N solutions appears in Figures 2b and 2d. In the TD99 model, the distribution of $J_{z}$ in the lower boundary is not uniform, and the P and N solutions recover this feature. The black S-shaped field lines of the bald patch are also {approximately} recovered in the P and N solutions. The self-consistent solution does not reproduce the bald patch field lines (Figure 2c). For the self-consistent solution there is still a central core to the flux rope (green field lines in Figure 2c), {but its location is changed \textit{cf.} the analytic solution.}
%
%
\begin{figure}
\centering
\includegraphics[width=1\textwidth]{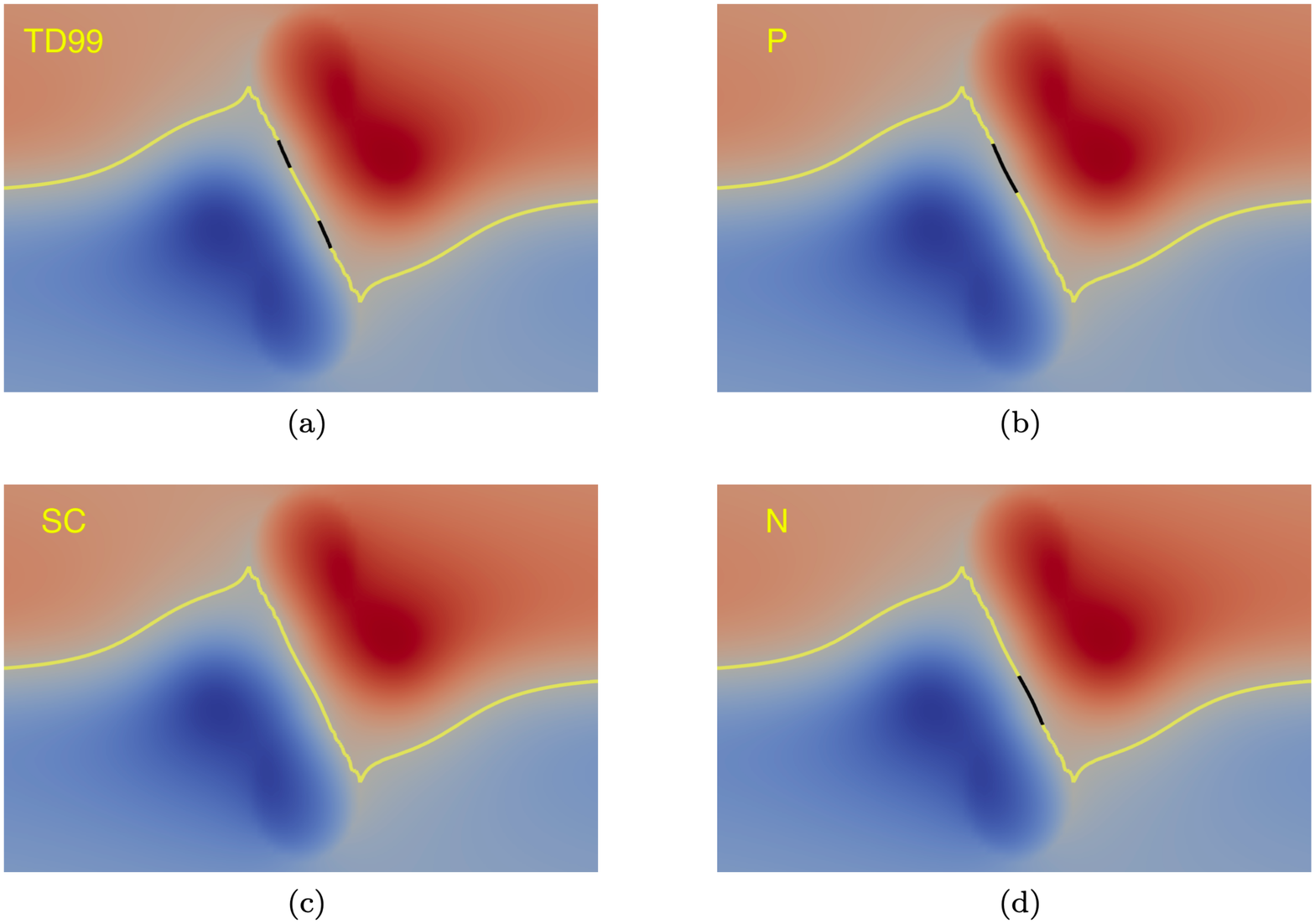}
\caption{Boundary values for Case 2 ($N_{\textrm{t}}=3.2$). The red and blue colours show regions in the lower boundary with $B_{z} > 0$ and $B_{z} < 0$ respectively. The yellow line corresponds to the neutral line and the black portions of the neutral line indicate  bald patches. The bald patch locations are determined using Equation \ref{bpequation}. The panels are arranged as follows: (a) analytic test field; (b) P solution; (c) self-consistent solution; (d) N solution.}
\end{figure}

\noindent Figure 3 shows the results for Case 5, with a surface twist number $N_{\textrm{t}} =8.8$. In this case the TD99 test field has strongly curved bald patch field lines, as shown in Figure 3a. The field lines near the centre of the flux rope (green) are approximately reproduced in each reconstruction. The bald patch field lines are reconstructed in the P and N solutions (black curves in Figures 3b and 3d respectively), but the results do not closely match the shape appearing in Figure 3a. As in Cases 2 and 3, no bald patch field lines are present in the self-consistent solution.
\\
\\
Figure 3 shows that the patterns of current density $J_{z}$ in the lower boundary for Case 5. {The results are similar to those for Cases 2 and 3, but the increased twist has clearly reduced the accuracy of the reconstructions.}
\\
\subsection{Locations of Bald Patches}\label{32} 
\begin{figure}
\centering
\includegraphics[width=1\textwidth]{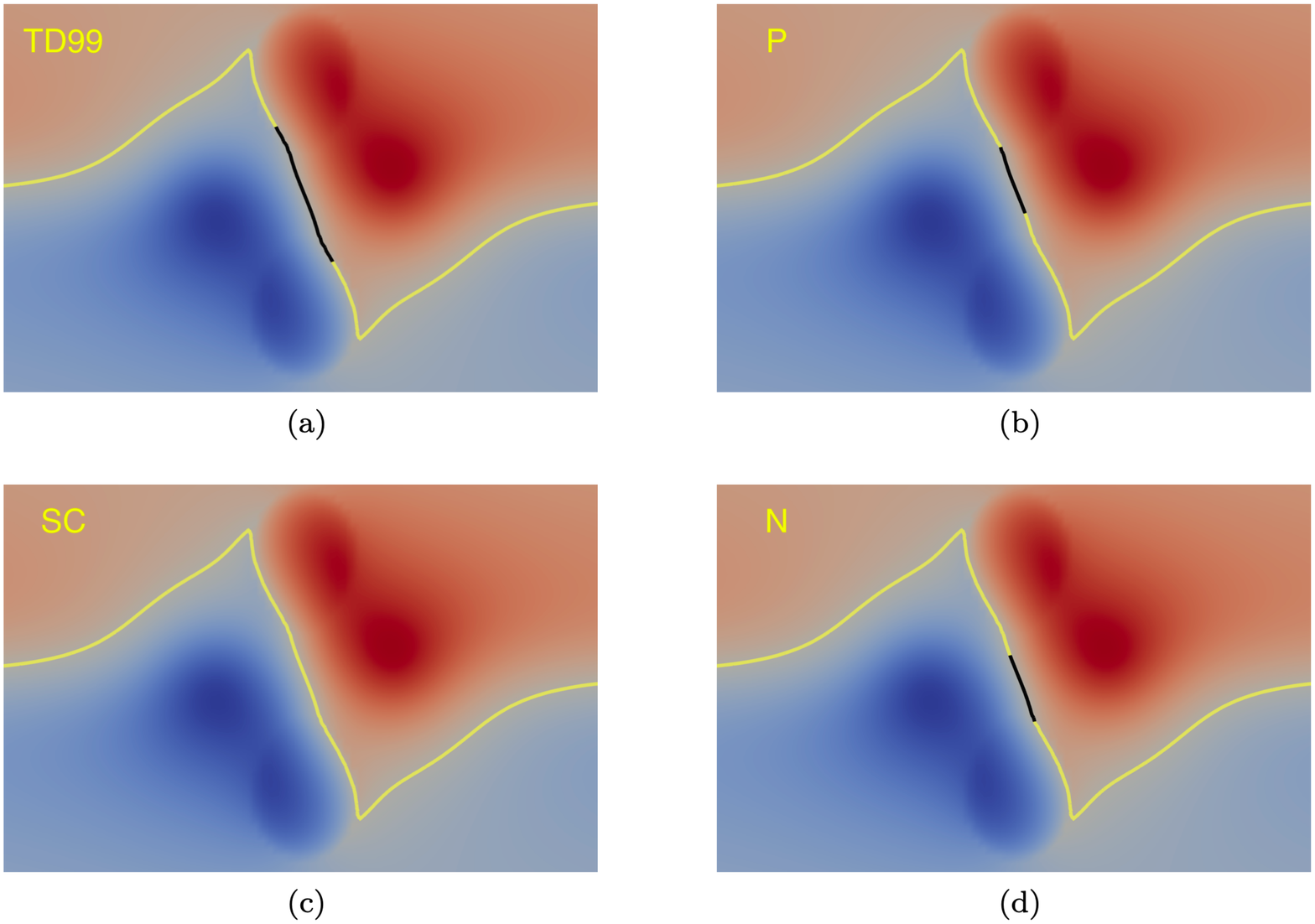}
\caption{Boundary values for Case 3 ($N_{\textrm{t}}=4.4$). The layout is the same as in Figure 4.}
\end{figure}

\noindent The magnetic fields of Cases 2, 3 and 5 have increasing twist ($N_{\textrm{t}}= 3.2 \;, N_{\textrm{t}}= 4.4 \; \textrm{and} \; N_{\textrm{t}}= 8.8$ respectively). These test fields include bald patches. 
\\
\\
Figure 4 shows the location of bald patches for Case 2. The neutral line is shown as a yellow line and the bald patch is a black segment along the neutral line. The bald patch location is calculated using Equation \ref{bpequation}. Both the P and N solutions recover a bald patch as shown in Figures 4b and 4d respectively. The test field (Figure 4a) contains a bifurcated bald patch, but the P and N solutions do not recover the bifurcation -- they contain a single bald patch, in each case. The self-consistent solution is not able to recover a bald patch as shown in Figure 4c. The averaging process in  the self-consistency procedure leads to the loss of the bald patch, this effect is discussed in Section 3.4.
\\
\\
Figure 5 shows the bald patches for Case 3. Again, both the P and N solutions recover a bald patch as shown in Figures 5b and 5d respectively, but the length and location of the bald patch is changed. The self-consistent solution is not able to recover a bald patch as shown in Figure 5c. 
\\
\\
Figure 6 shows the results for Case 5. The bald patches in the P and N solutions (Figures 6b and 6d) are less extended compared to the TD99 test field in Figure 6a. The self-consistent solution does not include a bald patch.

\begin{figure}
\centering
\includegraphics[width=1\textwidth]{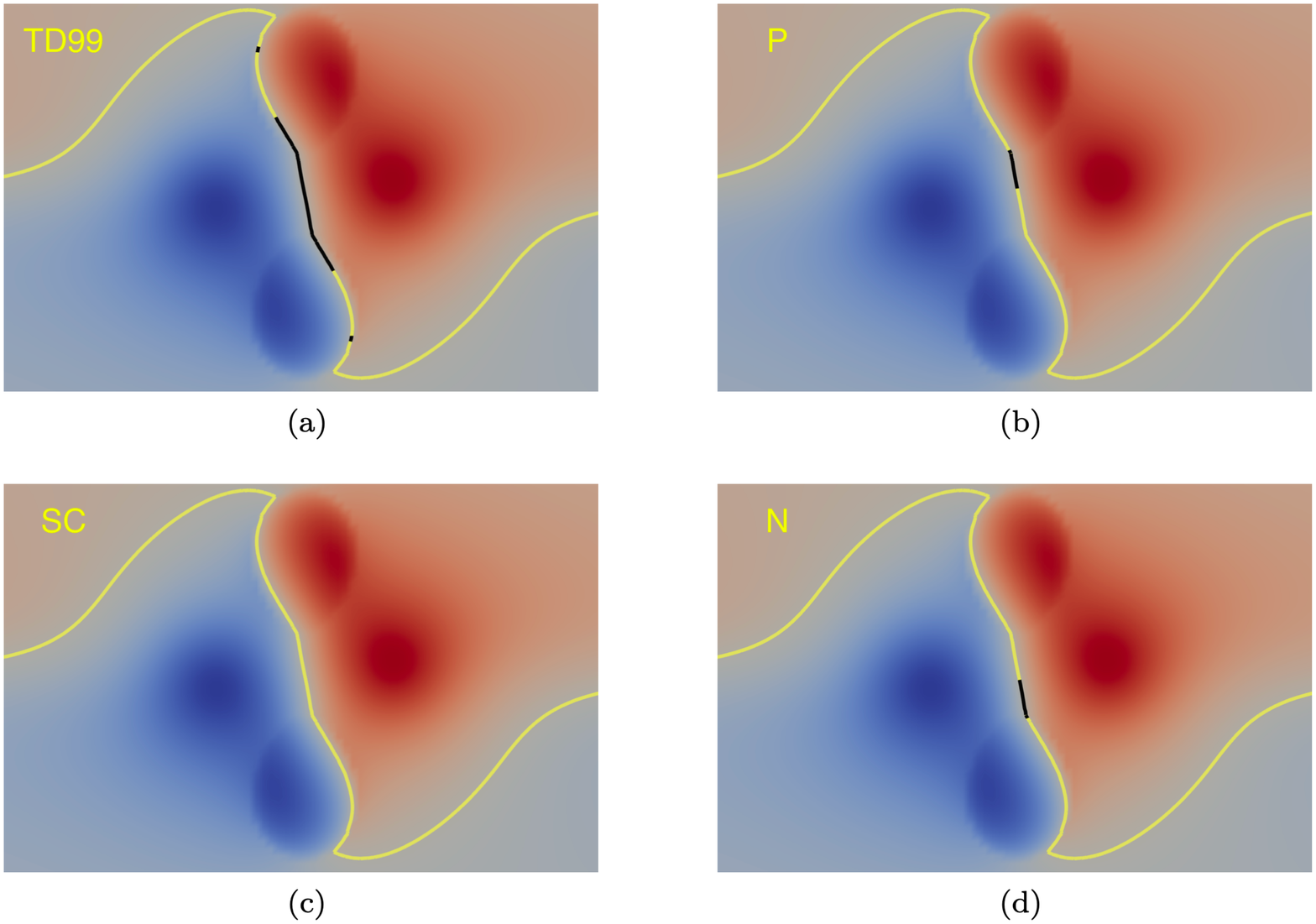}
\caption{Boundary values for Case 5 ($N_{\textrm{t}}=8.8$). The layout is the same as in Figure 4.}
\end{figure}

\subsection{Figures of Merit}\label{33} 
 Table 2 shows the metrics given by Equations \ref{em1} - \ref{emlast} calculated for the P, N and self-consistent solutions for the five sets of parameters listed in Table 1. The P and N solutions have the same metrics in every case because of the symmetry in the boundary conditions for the TD99 fields, so we combine these values in Table 2.
 
%
\begin{table}
\centering
\begin{tabular}{l*{7}{c}r}
\small{Field} & \small{Pol} & $N_{\textrm{t}}$ &  $C_{\textrm{vec}} $ & $C_{\textrm{cs}} $ & $E_{m}' $ & $E_{n}' $ & $ \epsilon$ \\
\hline
Case 1& P, N & 2.0 & 0.9986 & 0.9993 &0.9630 & 0.9604 & 0.9598  \\
\hline
Case 2 &P, N & 3.2 & 0.9957 & 0.9985 & 0.9546 & 0.9440 & 0.9401  \\
\hline
Case 3 & P, N & 4.4 & 0.9900 & 0.9971 &0.9435 & 0.9218 & 0.9109  \\
\hline 
Case 4 &P, N & 8.0 & 0.9509 & 0.9881 &0.8776 & 0.8104 & 0.7969  \\
\hline
Case 5 &P, N & 8.8 & 0.9485 & 0.9877 &0.8752 & 0.8048 & 0.7900  \\
\hline
\hline
Case 1 &SC & 2.0 & 0.9991 & 0.9995 &0.9642 & 0.9624 & 0.9574  \\
\hline
Case 2 &SC & 3.2 & 0.9959 & 0.9987 &0.9540 & 0.9423 & 0.9316  \\
\hline
Case 3 &SC & 4.4 & 0.9897 & 0.9973 &0.9416 & 0.9170 & 0.8963 \\
\hline
Case 4 &SC & 8.0 & 0.9401 & 0.9865 &0.8520 & 0.7629 & 0.7389  \\
\hline
Case 5 &SC & 8.8 & 0.9368 & 0.9860 &0.8470 & 0.7529 & 0.7290  \\
\hline
\end{tabular}
\caption{Comparison of the CFIT reconstructions with the TD99 test field using the metrics given by Equations \ref{em1} - \ref{emlast}. The corresponding TD99 parameters appear in Table 1. This table presents the error metrics for the P, N and self-consistent solutions, for all five cases.}
\end{table}

\noindent Table 2 indicates that the metrics and hence the quality of the reconstructions decrease with increasing twist. The metrics are not identically one for low twist because the TD99 field is only approximately force-free. 
\\
\\
The results are comparable to those reported for previous reconstructions (\textit{e.g.}  Wiegelmann \textit{et al.} 2006; Valori \textit{et al.} 2010b). It is not possible to make a one-to-one comparison because of the differences in the fields used, and the dependence of the results on the choice of the analysis volume.
\\
\\
The metrics indicate that the P and N solutions give better results than the self-consistent solutions, as already indicated by the field lines traces shown in Figures 1, 2 and 3. A possible explanation as to why the self-consistent method does not recover bald patches is that these features sensitively depend on the distributions of positive and negative $J_{z}$ in the lower boundary. The P and N solutions better preserve the distributions of $J_{z}$ in the lower boundary as shown in Figures 1, 2 and 3. The self-consistency procedure involves taking averages of values of $\alpha$ in the P and N solutions, over multiple cycles. In application to boundary conditions which are not exactly force free, this leads to a spreading of the current distribution. It is likely that this effect leads to the loss of the bald patches shown in Figures 1, 2 and 3. 

\section{Conclusion}\label{s:conclusions} 
In this paper we investigate the ability of a nonlinear force-free field code (CFIT) to reconstruct the Titov-D{\'e}moulin field \citep{Titov_demoulin_99} from its {boundary conditions at the lower boundary}. We explore the dependence of the results on increasing surface twist number $N_{\textrm{t}}$.  The results of five cases having surface twist numbers ranging from $N_{t} = 2$ to $N_{t} = 8.8$ are analysed. Inspection of field line traces and of the distribution of {the} current density at the lower boundary is used to identify the successes and limitations of CFIT when reconstructing twisted magnetic fields. The reconstructions recover important features of the fields, in particular bald patches, up to $N_{t} =8.8$, although the accuracy of the results based on visual comparison and the metrics of success decline with increasing surface twist number. Bald patches are recovered only for the P and N solutions \textit{i.e.} solutions using boundary conditions on $\alpha$ over only one polarity of the field in the lower boundary. Bald patches were not {recovered} for the self-consistent solutions, which use values of $\alpha$ over both polarities. Where bald patches were reconstructed, their sizes and locations did not exactly replicate those in the TD99 solutions.
 \\
 \\
Our investigation extends previous reconstructions of the TD99 model by using the original TD99 field values, without MHD relaxation, by using a Grad-Rubin approach (the CFIT code) that allows one to calculate of P and N solutions as well as a self-consistent solution, by fixing the geometric parameters of the TD99 field, and by including cases with high surface twist numbers.
\\
\\
The main results of this paper are summarised below.
\begin{itemize}
\item {Reconstructions are achieved using the Grad-Rubin method (CFIT) applied to approximately force-free boundary conditions \textit{i.e.} the boundary conditions for the original TD99 field. 
}

\vspace{3.5mm}

\item {The CFIT code can approximately reconstruct the TD99 field for a range of flux rope surface twist numbers. The flux-rope system, in particular the field lines near the centre of the flux rope, are approximately reproduced. A high-twist limit to our reconstructions is identified as Case 5 ($N_{\textrm{t}} \approx 8.8$). The figures of merit show a decline of the quality of the reconstructions with increasing $N_{\textrm{t}}$.}

\vspace{3.5mm}

\item {Bald patches in the TD99 fields are recovered using the P and N solutions, although their locations are changed with respect to the TD99 field. Bald patches are not recovered with the self-consistency procedure. }

\vspace{3.5mm}

\item {The self-consistency procedure does not improve the reconstruction of the TD99 field compared with the P/N solutions.}

\end{itemize}

\vspace{3mm}

\noindent The reconstructions considered in this paper use specific choices of parameters in the TD99 model, {and we use specific reconstruction methods (\textit{e.g.} the CFIT code, our method of constructing values of $\alpha$, \textit{etc.}).}  However, the results are indicative of our ability to reconstruct coronal magnetic fields from observational data, and in particular to recover {topological structures of interest.} 

\section{Acknowledgements}
\noindent VD is supported by the Australian Research Training Program. VD thanks Donald Melrose for helpful comments and suggestions on the manuscript. This work was funded in part by an Australian Research Council Discovery Project (DP180102408). We thank an anonymous referee for their work.

\newpage


\addcontentsline{toc}{section}{Bibliography}
\bibliographystyle{aasjournal}
	\bibliography{bibliography}

\end{document}